\documentclass[12pt, preprint]{aastex}

\usepackage{amsmath}
\usepackage{natbib}

\newcommand{\etal}{{et al.}}            

\newcommand{\nh}{$N_{\rm H}$}

\newcommand{\XMM}{{\em XMM-Newton }}

\newcommand{\Ch}{{\em Chandra }}

\shorttitle{X_RAY OBSERVATIONS OF CEN A NUCLEUS}
\shortauthors{EVANS ET AL.}

\begin{document}

\title{CHANDRA AND XMM-NEWTON OBSERVATIONS OF THE NUCLEUS OF CENTAURUS A}
\author{D. A. Evans\altaffilmark{1}, R. P. Kraft\altaffilmark{2}, D. M. Worrall\altaffilmark{1}, M. J. Hardcastle\altaffilmark{1}, C. Jones\altaffilmark{2}, W. R. Forman\altaffilmark{2}, S. S. Murray\altaffilmark{2}}
\altaffiltext{1}{University of Bristol, Department of Physics, Tyndall Ave., Bristol BS8 1TL, UK}
\altaffiltext{2}{Harvard-Smithsonian Center for Astrophysics, 60 Garden St., Cambridge, MA 02138, USA}

\begin{abstract}

We present X-ray spectra of the nucleus of the nearby radio galaxy Centaurus A from observations with the \XMM EPIC CCD cameras (two exposures separated by 12 months) and the \Ch HETGS. For the first time in an FRI type galaxy, we resolve fluorescent K$\alpha$ emission from cold, neutral, or near-neutral iron at 6.4 keV, with an rms line width of $\sim$20 eV. The Fe line parameters observed are consistent with fluorescent emission from material at a large distance from the active galactic nucleus, either in the form of an absorber that nearly completely surrounds the central engine or a torus that lies predominantly out of the line of sight.  Unresolved emission lines from neutral Si~K$\alpha$ at 1.74 keV and neutral S~K$\alpha$ at 2.30 keV are also detected. We find no evidence in the data for a previously reported 6.8 keV broadened Fe line. The continuum spectrum is well fitted with a combination of a heavily absorbed power-law component that we relate, using Bondi theory, to accretion phenomena in the form of a standard, geometrically thin, optically thick disk, and a second, less absorbed, power-law component that we associate with emission from the subparsec VLBI radio jet.

\end{abstract}

\keywords{galaxies: active - galaxies: individual (Centaurus A, NGC 5128) - galaxies: jets - X-rays: galaxies}

\section{INTRODUCTION}

Centaurus A (NGC 5128, Cen~A) is the nearest ($d \sim$ 3.4 Mpc, $1''\sim$ 17 pc) active galaxy to the Milky Way. It exhibits complex morphology on scales ranging from milliarcseconds to degrees on the sky, and shows a rich variety of structure across the electromagnetic spectrum. Optically, the host of Cen~A is an elliptical galaxy with a large bisecting dust lane, which is believed to be evidence for a recent merger. Resolved X-ray emission from the predominantly one-sided kiloparsec-scale radio jet has been detected in detail with \Ch (\citealt{kra00,kra02,har03}), and measurements of the X-ray hot gas environment of Cen~A using \Ch and \XMM have been reported by \cite{kar02} and \cite{kra03}.

The nuclear X-ray continuum spectrum of Cen~A is described by a highly absorbed power law, with a photon index $\Gamma$ $\sim$ 1.7 and absorbing column $\sim$10$^{23}$ atoms cm$^{-2}$ (e.g., \citealt{tuc73,mus78}). Two popular interpretations for the origin of the continuum are (1) the emission is associated with accretion phenomena in or near a disk (e.g., \citealt{tur97}) and (2) the observed emission is from synchrotron self-Compton upscattering of the subparsec scale radio jet to X-ray energies (e.g., \citealt{gri75,chi01}). The high absorbing column might, for example, originate from a dusty torus surrounding the accretion disk (e.g., \citealt{woz98}), or material that surrounds (4$\pi$ sr) the nucleus (e.g., \citealt{miy96}).

Previous X-ray satellites have detected unresolved line emission at $\sim$6.4 keV in Cen~A, corresponding to K$\alpha$ fluorescence from neutral or near-neutral states of iron (e.g., \citealt{mus78}). The detection of less prominent lines from neutral Mg, Si and S, has also been claimed (\citealt{sug97}). \emph{BeppoSAX} observations of Cen~A, carried out between 1997 and 2000, found evidence for two separate Fe K$\alpha$ lines: one cold and narrow, with a centroid energy of 6.4 keV, and the other broad and variable, with a centroid energy of 6.8 keV (\citealt{gra03}). The observations were used to argue in favor of a hot, geometrically thick, optically thin accretion flow surrounded by material with different opacities. However, it should be noted that the uncertainties associated with the parameters of the 6.8 keV line were considerable.

Nuclear X-ray spectra of other low-power radio galaxies are commonly described by a power-law continuum, absorbed by a wide range of column densities, together with one or more fluorescent emission lines. Typical examples are 3C66B (\citealt{har00}), M87 (\citealt{wil02}), NGC 6251 (\citealt{gli04}) and NGC 4261 (\citealt{gli03}). The physical origin of the continuum spectrum in these other low-power galaxies, as with Cen~A, is still uncertain. Some of the soft nuclear X-ray emission may be associated with emission from the base of a jet (\citealt{wor94,can99}), and hard components have been interpreted as either due to accretion processes (e.g., \citealt{tur97b,gli04}) or a jet. Bondi theory (\citealt{bon52}) may be applied to calculate the radiative efficiency, $\eta_{\rm Bondi}$, of an accreting supermassive black hole. If the accretion process has $\eta_{\rm Bondi}$ $\sim$ 1--10\% (e.g., \citealt{gli04}), it is likely that the accretion flow takes the form of a geometrically thin, optically thick disk (\citealt{sha73}). Alternatively, if the radiative efficiency is much less than the canonical value (e.g., \citealt{gli03}), the accretion flow process may be fundamentally different, and might take the form of a geometrically thick disk (e.g., \citealt{nar95}).

In this paper, we present the results from two \XMM EPIC observations and a \Ch HETGS observation of X-ray emission from the nucleus of Cen~A. For the first time in an FRI type galaxy, we resolve the fluorescent iron K$\alpha$ line emission using \Ch HETGS and show that the observed line energy, equivalent width, and variability properties are consistent with emission from a region distant from the central black hole. We present a detailed analysis of the X-ray continuum spectrum, and demonstrate that a more complex model than a single powerlaw is necessary to explain adequately the observed emission at low energies. We find that a second, less absorbed, soft power-law component can explain the low-energy emission and is likely associated with X-ray emission from the subparsec-scale radio jet. We apply Bondi accretion theory to show that the hard continuum can be interpreted as originating from an accretion flow in the form of a standard, geometrically thin disk.

The recessional velocity of Cen A corresponds to a redshift of $z$=0.0018. Because of Cen~A's proximity and peculiar velocity, however, this redshift is not indicative of its distance. The distance to Cen~A has been measured by a variety of techniques, and a range of values have been quoted in the literature.  The most recent measurement places it at 3.84 $\pm$ 0.35 Mpc (\citealt{rej04}), somewhat farther than the value 3.4 $\pm$ 0.15 Mpc claimed by \cite{isr98}.  We used this latter value in our five previous papers on X-ray observations of Cen~A, and for consistency we have adopted it in this paper.  None of our conclusions would be changed if we adopted the larger value; the only effect would be to increase all luminosities by $\sim$30\%.  All spectral fits include absorption through our Galaxy using $N_{\rm H, Gal}$ = 7.69$\times$10\(^{20}\) atoms cm$^{-2}$  (\citealt{dic90}). When comparing Cen A with other objects at cosmological distances, we use a cosmology in which $\Omega_m$ = 0.3 and $\Omega_\Lambda$ = 0.7, and we adopt H$_0$ = 70 km s$^{-1}$ Mpc$^{-1}$.

\section{OBSERVATIONS AND ANALYSIS METHODS}

\subsection{XMM-Newton EPIC}

Cen~A was observed twice with the \XMM EPIC instrument, on 2001 February 2 ($\sim$23 ks exposure) and 2002 February 6 ($\sim$13 ks exposure). In both observations the three EPIC CCD cameras (MOS1, MOS2, and pn) were operated with the medium optical blocking filter in place. The MOS cameras were operated in full-frame mode, and the pn camera in extended full-frame mode. The results presented here were obtained using the latest software and calibration files available from the \XMM project, SAS version 5.4.1. Calibrated event files were generated using the EMCHAIN and EPCHAIN scripts.

To check for intervals of high particle background, light curves were extracted for the whole field of view, excluding a $100''$ radius circle centered on the source. The events were filtered to include only those with PATTERN=0 and FLAG=0 attributes and for an energy range of 10--12 keV for the MOS cameras and 12--14 keV for the pn, where the particle background should dominate the emission. The light curves showed that there were no times of high background, which we defined following \cite{fil03} as more than 15 counts per 100 s time bin for the MOS1 and MOS2 cameras and more than 25 counts per 100 s time bin for the pn camera. For our analyses we measure background locally from large circles away from the source. A co-added MOS1, MOS2, and pn image of the second \XMM observation is shown in Figure~\ref{xmmcombined} with the area chosen for background subtraction marked.

The \XMM observations of the nucleus of Cen~A are heavily piled-up. We
demonstrate this using two independent methods.  First, we determined
the count rates in a 100$''$ source-centered circle with no PATTERN or
FLAG filtering. As a representative case, the count rates in the
second \XMM observation were found to be 4.03 and 15.37 counts s$^{-1}$ for the for the MOS1 and pn cameras, respectively.  The maximum recommended count rates to avoid significant pileup are 0.70 and 3 counts s$^{-1}$ for the MOS1 and pn cameras, respectively (\citealt{xmmuhb}). As a second independent estimate, we followed the prescription of \cite{mol03} to produce an image of diagonal bipixels in the MOS cameras (assigned PATTERN numbers 26--29), which are produced almost solely by the pileup of two single-pixel events. We computed the ratio of number of diagonal bipixel events within a source-centered circle of radius 20$''$ to that within an annulus of radii 20$''$ and 50$''$.  For the second \XMM observation, this ratio was found to be 32 $\pm$ 12.  The nominal ratio for unpiled up data is $\sim$2.4.  The large excess of diagonal bipixel events in the central 20$''$ demonstrates that this region is strongly affected by pileup. This method works only for the MOS cameras; for the pn camera, an image filtered with the FLAG = XMMEA\_0 parameter was produced for energies $\le$10 keV. This filtering parameter applies to events with a pattern that cannot be produced by a single photon i.e., those predominantly produced by pileup. This analysis shows for the pn camera that pileup is insignificant outside a source-centered circle of radius $20''$ (see Fig.~\ref{pn_xmmea0}).

In order to avoid complications due to pileup, we used an annulus for the spectral extraction region, thereby sampling the wings of the PSF and excluding the most heavily piled-up region. From the analysis described in the preceding paragraph, the inner radius of the annulus was chosen to be 20$''$. In determining the outer radius, there is a trade-off between the number of nuclear counts and the amount of underlying thermal galactic emission (not the focus of this investigation). We adopted 50$''$, as a further increase in outer extraction radius would mean that we begin to sample knot B of the northeast kiloparsec-scale jet, and point sources to the southwest. To further minimize the effect of pileup, the event tables were filtered to include only events with the FLAG = 0 and PATTERN = 0 parameters. Redistribution matrix files (RMFs) and ancillary response files (ARFs) were generated using the SAS tasks \emph{rmfgen} and \emph{arfgen}. The ARF-generation corrects for the energy-dependent fraction of the PSF counts included in the annulus, parameterized by a King model (\citealt{ghi01a,ghi01b}).

In our \XMM analysis, the $20''-50''$ annulus contains both extended
emission from the jet and hot gas and pointlike emission from the core in which we are interested here. To measure the spectrum of the nucleus alone, we used \emph{Chandra}/ACIS-I data (observation date 2000 May 17; \citealt{kra01}) to model spectrally the contribution from the extended components in the $20''-50''$ annulus. This emission was treated as an additional background in the \XMM data by subtracting it from the \XMM spectra. We used the appropriate ARFs for pointlike and extended emission in our analysis.

\subsection{Chandra HETGS}

The nucleus of Cen A was observed twice 2 weeks apart with the
{\emph{Chandra}} HETGS; first on 2001 May 9and again on 2001 May
21. The live times of the two observations were 46.8 and 51.5 ks, respectively.  Both observations were roll-constrained so that the jet was placed along the cross-dispersion direction to avoid contamination of the dispersed spectrum of the nucleus.  The data were examined for periods of high background, and none were found. The nuclear spectra were extracted and grating ARFs and RMFs were created for both the HEG and the MEG data using the CIAO software package. Initially the two datasets were analyzed separately to search for temporal variability in either line or continuum features on the scale of weeks separating the two observations. No statistically significant differences were found, so the two datasets were combined, yielding a total observation time of 98.3 ks. Background was determined from a region offset from the dispersed spectra beyond the extraction region.  Contamination of the nuclear spectrum from the other X-ray emitting components present in Cen A, including the hot gas and the X-ray binary population, was generally negligible for two reasons.  First, the nuclear flux dominates the X-ray emission from the galaxy, and its spectrum is heavily absorbed ($N_{\rm H} >$ 10$^{23}$ atoms cm$^{-2}$), whereas the spectra of the other components are much softer and orders of magnitude less luminous. Second, the event processing/order sorting algorithm used to extract grating spectra efficiently removed the contribution from these other components unless they are located within a few arcseconds of the core.

\section{THE X-RAY SPECTRUM}

\subsection{Continuum Spectrum}

\subsubsection{XMM-Newton EPIC}

The \XMM spectral fitting was performed jointly for the MOS1, MOS2, and pn cameras for an energy range of 1.5--8 keV for each of the two observations, so as to avoid the low energies where thermal emission dominates the spectrum, as indicated by {\emph{Chandra}/ACIS imaging data. The net counts (1.5--8 keV) in our spectral fits to data from the annular extraction regions for the first and second observations, respectively, are MOS1: 11,773, 6,953; MOS2: 11,726, 7,086; pn: 21,251, 10,811. The data were grouped to 100 counts per PHA bin.

We initially modeled the nuclear continuum spectrum with a single,
heavily-absorbed power-law. The normalizations of all spectral
components were kept free for each camera. An Fe K$\alpha$ emission
line at $\sim$6.4 keV was included and modeled with a Gaussian
function, and is discussed in detail in \S 3.2. The fit was poor; for
example, for the second observation, $\chi$\(^{2}\) = 393 for 207 dof,
with large contributions to $\chi$\(^{2}\) at low energies (see
Fig. ~\ref{xmm_1pl}). A substantial improvement in the fit was
achieved by adding a second power-law component to the model (for
example, for the second \XMM observation, $\chi$\(^{2}\) = 211 for 203
dof). The spectral parameters for the second power law are poorly
constrained, so we decided to fix its photon index to 2. The best-fit
spectral parameters for this model, together with 90\% confidence
errors for one interesting parameter ($\chi$\(^{2}\)$_{\rm min}$ +
2.7), are given in Table~\ref{2pow_fitted} for both \XMM
observations. An unfolded spectrum for each \XMM observation is given in Figure~\ref{xmm_2pl_ufs}.

When estimating the parameters of the second power law we must take into account the statistical uncertainties in the model parameters for the resolved emission from jet and gas measured with \emph{Chandra}. This was achieved by using maximum and minimum amounts of this emission, as described by 90\% confidence limits. Using these upper and lower limits, we plot 90\% confidence contours for the second power law for the MOS1 and pn cameras in Figure~\ref{eee_obs1_90} for the first \XMM observation. Results for the MOS2 are consistent with those for the MOS1, as are those for the first and second observations. We also estimate how systematic errors in the \Ch annulus subtraction might affect the results of the second power law measured with \emph{XMM-Newton}. A systematic error approaching a factor of 2 was required before any significant change in second power law properties was measured, within the uncertainties.

We independently confirmed the existence of the second power law
component by analyzing data from the \Ch ACIS-I frame transfer
streak. As can be seen from the \XMM best-fitting unfolded spectrum in
Figure~\ref{xmm_2pl_ufs}, in the energy range 1.7--2.2 keV there is
only a small contribution to the flux from the heavily absorbed
primary power law. In this energy range, we measure a second power law
flux of $\sim$(8 $\pm$ 4)$\times$10\(^{-13}\) ergs cm$^{-2}$ s$^{-1}$
(90\% confidence). The ACIS frame transfer streak in the same energy
range should contain a similar flux and, crucially, does not contain a
contribution from the kiloparsec-scale jet (which, if modelled
incorrectly, could have affected the measured \XMM core
flux). Following \cite{mar04}, we determined the number of counts in
the frame transfer streak using a rectangular extraction region
surrounding the streak, using background determined from an adjacent
region devoid of point sources. The rectangle was wide enough that $>$
95\% of the source flux in the streak at any point was included. To
determine the number of source events that occured in the frame
transfer streak, we scaled the number of counts measured from the core
spectrum flux by the ratio of the pixel exposure time to the frame transfer time. We compared the measured 1.7--2.2 keV count rate in the ACIS-I frame transfer streak [(1.34 $\pm$ 0.19)$\times$10\(^{-3}\) counts s$^{-1}$] with that expected for an absorbed power law with a flux of (8 $\pm$ 4)$\times$10\(^{-13}\) ergs cm$^{-2}$ s$^{-1}$ [(1.23 $\pm$ 0.62)$\times$10\(^{-3}\) counts s$^{-1}$], and found the two to be consistent.

The normalization of the pn camera was lower than that of the MOS cameras for the second power-law component in both \XMM observations. The trend for the pn normalization to be low was also apparent (although not statistically significant) in the first power-law component (see Table~\ref{2pow_fitted}). We attribute this discrepancy to remaining uncertainties in the calibration between the MOS and pn cameras.

Since we cannot rule out the possibility that the two power-law indices are the same, we tested a partial covering model in which a certain fraction of the emission is covered by an absorbing column $N_{\rm H, 1}$  + $N_{\rm H, 2}$ and the remainder by an absorbing column $N_{\rm H, 2}$. A Gaussian Fe emission line was included in the fit. This yielded an acceptable fit, for example, for the second \XMM observation, $\chi$\(^{2}\) = 215 for 205 dof. The counts spectrum and best-fitting model, together with contributions to $\chi$\(^{2}\) is shown in Figure~\ref{xmm_zpc_obs2} for the pn camera. These best-fitting spectral parameters for both \XMM observations are given in Table~\ref{xmm_zpc_obs2_table}.

\subsubsection{Chandra HETGS}

The \Ch spectral fitting was performed jointly for the high-energy
grating (HEG) and medium-energy grating (MEG), over an energy range of 1.8--7 keV. The net counts in this energy range are 65,804 in the MEG and 68,400 in the HEG. The data were grouped to 100 counts per bin. A model containing a single, absorbed power law and an Fe K$\alpha$ line yielded an acceptable fit ($\chi$\(^{2}\) = 469  for 528 dof). An initial analysis showed that the HEG and MEG normalizations agreed, so the normalizations of all spectral components were kept linked across the gratings. A detailed analysis of the line emission features is given in \S 3.2. 

Guided by the inadequacy of a single, absorbed power-law fit to the \XMM data, we explored the possible range of parameter values that a second power-law component could take in the \Ch data, despite the relative insensitivity compared to \XMM in the energy range where this component is prominent. An improvement in the fit ($\Delta$$\chi$\(^{2}\) = 4.31 for one additional parameter) was obtained by adding such a component with a frozen photon index of 2 and an absorption of 3.8$\times$10\(^{22}\) atoms cm$^{-2}$, corresponding to the parameters of the second power-law in \S 3.1.1. The 90\% confidence upper limit of the 1 keV normalization is $\sim$3$\times$10\(^{-3}\) photons cm$^{-2}$ s$^{-1}$ keV$^{-1}$, which is consistent with the \XMM data. A counts spectrum is shown in Figure~\ref{chandra_2pl_counts}. The best-fit spectral parameters, together with 90\% confidence errors for one interesting parameter ($\chi$\(^{2}\)$_{\rm min}$ + 2.7), are given in Table~\ref{2pow_fitted}. This shows that the \Ch and \XMM spectra are consistent.

\subsection{Emission Lines}

\subsubsection{Chandra HETGS}

We used the \Ch HETGS data at full resolution to study narrow emission line features. A heavily absorbed power law was jointly fitted to each data set for energy ranges of 4.0--6.0 keV and 7.2--8.0 keV, and the interpolated continuum subtracted from the data in the energy range 6.0--7.2 keV. The continuum parameters were consistent with those found in \S 3.1.

Neutral Fe K$\alpha$ and K$\beta$ emission lines were fitted to the data and checked using an $F$-test for their statistical significance. The ratio of the intensities of the K$\alpha$ and K$\beta$ lines was fixed to 150/17 when considering models containing both lines. In order to correctly evaluate any Fe K$\alpha$ line broadening, we modeled this emission as a doublet, with the energy difference between the K$\alpha$$_1$ and K$\alpha$$_2$ lines fixed at 13 eV (\citealt{bea67}), the intensity of the K$\alpha$$_2$ line half of the K$\alpha$$_1$ line, and made the line widths vary with each other. The addition of an Fe K$\alpha$ line was highly significant ($\Delta$$\chi$\(^{2}\) = 90 for three additional parameters). The Fe K$\alpha$$_1$ line had a measured centroid of 6.404 $\pm$ 0.006 keV and was found to be broadened, with a width of 20 $\pm$ 10 eV (90\% confidence). The addition of an Fe K$\beta$ line was found to be statistically insignificant.

We also attempted to detect the broad 6.8 keV ionized iron line in the data, as claimed by \cite{gra03} based on \emph{BeppoSAX} observations. We found an insignificant improvement in the fit ($\Delta$$\chi$\(^{2}\) = 0.38 for two additional parameters) with the inclusion of this line. Further, the best-fitting intensity of this line was consistent with zero. Parameter values are given in Table~\ref{line_parameters}, and a counts spectrum showing the contributions of the Fe K$\alpha$$_1$ and K$\alpha$$_2$ lines is presented in Figure~\ref{chandra_iron_counts}.

In addition, we searched for low-energy emission lines using MEG-1 data grouped to 15 counts per energy bin. The inclusion of a Si K$\alpha$ line at 1.74 keV was significant at $>$ 99.9\% on an $F$-test. We also found a significant improvement in the fit upon adding a S K$\alpha$ line at 2.30 keV. A counts spectrum showing the contributions of Si and S is shown in Figure~\ref{chandra_Si_S}, and parameter values are given in Table~\ref{line_parameters}.

\subsubsection{XMM-Newton EPIC}

The \XMM data were used to study the Fe line emission, but not
contributions from Si and S, because of the lower energy resolution of
\XMM EPIC compared to \Ch HETGS. We examined data from the \XMM RGS
camera, but found that they had an insufficient signal-to-noise ratio
to perform a useful analysis. We used the pn camera only for the \XMM
analysis, because of its larger effective area and better statistics as compared with either the MOS1 or MOS2 cameras. The method of continuum subtraction and analysis described in \S 3.2.1 was followed throughout. The residuals for the two \XMM observations are shown overlaid in Figure~\ref{xmm_iron_residuals}. The line widths of all emission lines models were frozen to be narrow in \emph{XSPEC}, with $\sigma$ = 10 eV (small compared with the $\sim$130 eV FWHM resolution of the EPIC pn camera at 6.4 keV). 

The inclusion of an Fe K$\alpha$ line was statistically significant at
greater than 99.9\% on an $F$-test. The 90\% confidence lower limit on the line centroids for each observation are 6.408 keV and 6.413 keV, marginally inconsistent with the centroid of 6.404 keV, measured with \Ch HETGS. This may be interpreted as due to a systematic error in the calibration of the energy scale of the pn camera, causing the line centroid to be apparently shifted to a slightly higher energy. This effect has been mentioned elsewhere (e.g., \citealt{wor03}). The line energy, line width, equivalent width, and unabsorbed flux of the Fe K$\alpha$ line for both \XMM observations are given in Table~\ref{line_parameters}. There is no statistically significant evidence for an Fe K$\beta$ line in the data.

We also searched for a broad, ionized emission line in the data by following the method used in the \Ch analysis. We found no statistically significant evidence for a blue tail to the Fe K$\alpha$ emission between an observed energy of 6.5 and 7.2 keV (see Fig.~\ref{xmm_iron_residuals}), unlike that reported in \cite{gra03}. Following the method of \cite{gra03}, we found no evidence in the data for a $\sim$6.8 keV line and, when the parameters of the line were set to equal those in \cite{gra03}, we found a worse fit ($\Delta$$\chi$\(^{2}\) = 4.73). When the intensity of the line was set to the 90\% confidence lower limit in \cite{gra03}, the fit was still worse ($\Delta$$\chi$\(^{2}\) = 1.22).

\section{SOURCE VARIABILITY}

X-ray light curves were created using the pn camera for each \XMM
observation using a source-centered circular extraction region with a
radius 30$''$. Background subtraction was applied using a
source-centered annular extraction region of inner radius 400$''$ and
outer radius 600$''$. To examine any variability of the hard, absorbed
power law, an energy range of 3--10 keV was used. An energy range of 0.5--2.5 keV was also used to search for variability of the softer, second power-law component. Only events with FLAG = 0 and PATTERN = 0 parameters were selected.

To search for any intra-observation variability, the data were binned
in time intervals of 500, 1000, and 2,000 seconds, and a $\chi$$^{2}$
analysis used to test the null hypothesis. We found no significant
intra-observation variability, with typical values of $P_{\chi^{2}}$
$>$ 50\%. In addition, we calculated the fluxes from the \XMM and \Ch
count rates in the energy band 4--7 keV: i.e., where the
heavily-absorbed power law is likely to dominate and where there is
reasonable effective area in both instruments. The \XMM fluxes were
calculated from 20$''$ to 50$''$ source-centered annular extraction
regions in the pn camera data for both observations. The \Ch fluxes
were found using the HEG+1 data. The results are given in
Table~\ref{4to7var}. Because of the high count rate, the statistical errors were small. We found an increase in the 4--7 keV unabsorbed flux of $\sim$19\% between the two \XMM observations, and a $\sim$57\% increase between the first \XMM observation and the \Ch observation, similar in magnitude to the 60\% variations previously reported (\citealt{ben01}). We also note that the 0.5--2.5 keV count rate increased by $\sim$48\% between the two \XMM observations. Finally, we found that the Fe K$\alpha$ flux did not appear to correlate with the 4--7 keV continuum flux, as shown in Figure~\ref{iron_cont_corr}, and that any Fe K$\alpha$ flux variability was not statistically significant.

\section{INTERPRETATION}

\subsection{Continuum}

The nuclear spectrum of Cen~A is well described by a two-component
power-law model, with emission lines from Si, S, and Fe. This result
is consistent with two distinct models of the active galactic nucleus geometry. The first (\citealt{tur97}) is that $\sim$90\% of the nucleus of Cen~A is covered by an absorbing column of $\sim$(1.1--1.4) $\times$10\(^{23}\) atoms cm$^{-2}$ and $\sim$10\% by a column of $\sim$(4.4--4.5) $\times$10\(^{22}\) atoms cm$^{-2}$. This partial covering model gives the required low energy flux. A second possible interpretation of the data is that we are observing unresolved X-ray emission from two power-law components, each with some degree of intrinsic absorption. The second, less-absorbed component may be consistent with arising from the subparsec radio VLBI jet. High-resolution VLBI observations of Cen~A (\citealt{tin98}) have revealed a variety of radio structures, including a compact, self-absorbed core with a highly inverted spectrum and a subparsec-scale jet, extending to $\sim$40 mas (0.7 pc). Emission from the core does not dominate over that from the jet in the VLBI observations, even at 8.4 GHz, where it is most prominent. At 4.8 GHz the flux is dominated by emission from the jet. We also note from \cite{tin98} that radio variability by a factor of $\sim$3 occurs in the self-absorbed core, and $\sim$70\% in the subparsec-scale jet, both on timescales of years.

Using a normalization of $\sim$5$\times$10\(^{-3}\) photons cm$^{-2}$ s$^{-1}$ keV$^{-1}$ (i.e. comparable to both the \XMM and \Ch spectral fits) we calculate the flux density ratio of the 1 keV second power law X-ray emission and the parsec-scale jet seen in VLBI observations (\citealt{tin98}). We perform a similar calculation with the kiloparsec-scale jet, using {\emph{Chandra}}/ACIS-I observations (\citealt{kra02}) and 8.4 GHz VLA observations (\citealt{har03}) in a source-centered $20''$--$50''$ annulus. The fluxes of each component, together with their ratios are given in Table~\ref{fluxratios}. We find the X-ray to radio flux ratio for the VLBI jet and the second X-ray power-law component to be similar to that from the kiloparsec-scale jet emission detected with the \Ch CCDs and the VLA radio observations, which might imply a synchrotron origin for this unresolved X-ray emission.

Previous studies of the X-ray and radio flux properties of the active nuclei of low-power radio galaxies (\citealt{fab84,wor94,can99,har99}) have shown a correlation between the luminosities of soft, unresolved X-ray emission and 5 GHz core radio emission. Figure~\ref{b2_cena_corr_lumin} shows that the measured X-ray luminosity of the second power law detected in Cen~A lies on or near the trendline established for the large B2 sample of radio galaxy cores (\citealt{can99}). This is consistent with the hypothesis that the observed low-energy X-ray emission originates from the subparsec-scale VLBI jet.

\subsection{Emission Lines}

The measured line energy and width of the Fe K$\alpha$ line, together
with its lack of response to continuum variations implies an origin in
cool, neutral or near-neutral material far from the central black
hole. This hypothesis is supported by the detection of unresolved
emission lines from neutral Si and S. The measured Fe K$\alpha$ line
width and the estimated black hole mass of 2$\times$10\(^{8}\)
$M_\odot$ (\citealt{mar01}) give an Fe emission radius of of 0.1 pc (6
mas) using Keplerian arguments. Note that this value of $M_{\rm BH}$
is higher than the values presented in \cite{isr98}. Two models have
previously been proposed to explain the possible geometry of the
fluorescent region being illuminated by a primary power-law
continuum. In the first model (\citealt{miy96}), the emission arises
from a cold cloud with an absorbing column of $\sim$10\(^{23}\) atoms
cm$^{-2}$ that entirely surrounds the nucleus. The predicted
equivalent width of $\sim$60--70 eV in this model is roughly
consistent with that measured here. In the second model
(\citealt{woz98}), the central engine is surrounded by a structure
with an intrinsic absorption of $\gg$ 10\(^{23}\) atoms cm$^{-2}$ that
lies predominantly outside of the line of sight, such as a torus. The
data presented here cannot distinguish between the two models,
although it seems that line emission originating predominantly from
the inner regions of an accretion disk is unlikely. In addition, the
lack of observed line emission at higher energies ($\sim$6.7 keV)
disfavors a model in which ionized species of Fe in an accretion disk
close to the central black hole emit fluorescent lines. However, our
results are consistent with a model in which a standard accretion disk
emitting a fluorescent line at 6.4 keV is truncated and replaced by a
hot radiatively inefficient accretion flow in the inner regions near
the central engine, as discussed for NGC 4258 by, for example, \cite{gam99} and \cite{rey00}.

\subsection{Nature of the accretion flow}

We estimate the rate, $\dot{M}_{\rm Bondi}$, at which the central black hole in Cen~A gravitationally captures gas by applying Bondi accretion theory (\citealt{bon52}), and calculate the Bondi radiation efficiency, $\eta_{\rm Bondi}$, by comparing the measured X-ray luminosity, $L_{\rm X}$, with the Bondi luminosity, $\dot{M}_{\rm Bondi} c$\(^{2}\). The Bondi accretion rate may be written as

\begin{equation}
\dot{M}_{\rm Bondi}=4{\pi}R^2_{\rm A}\rho{\rm _A}c_{\rm s}, \label{bondi1}
\end{equation}
\\where the accretion radius, $R_{\rm A}$, is given by $R_{\rm A}{\simeq}GM/{c^2_{\rm s}}$, with $\rho_{\rm A}$ the density of gas at $R_{\rm A}$, and $c_{\rm s}$ the sound speed.

We follow previous work in assuming that the gas supply for the
central black hole is the X-ray--emitting hot gas, with a typical
temperature of 0.5 keV. It should be noted that there are large
amounts of cold, dense gas in the center of Cen~A, so the Bondi
accretion rate calculated here is a lower limit. We take the black
hole mass to be 2$\times$10\(^{8}\) M$_\odot$, based on Very large Telescope infrared spectroscopy measurements of Cen~A (\citealt{mar01}). This gives a maximum accretion radius of 26 pc from the central black hole. At this radius, we take the gas density to be the interpolated central value of 3.7$\times$10\(^{-2}\) cm$^{-3}$ (\citealt{kra03}), which is uncertain by factors of a few, and find $\dot{M}_{\rm Bondi}$ = 6.41$\times$10\(^{-4}\) M$_\odot$ yr$^{-1}$.

The 2--10 keV luminosity found with these observations of Cen~A is $\sim$5$\times$10\(^{41}\) ergs s$^{-1}$. If Cen~A accretes at the Bondi rate, we compare this measured X-ray luminosity with $\dot{M}_{\rm Bondi} c$\(^{2}\), to find an upper limit to the Bondi efficiency, $\eta_{\rm Bondi}$ $\sim$0.2\%.  In Table~\ref{bondi}, we compare $\eta_{\rm Bondi}$ for Cen~A with other galaxies (\citealt{qua03,bag03,loe01,dimatt03,gli03,gli04}).

From Table~\ref{bondi}, Cen~A has an efficiency $\eta_{\rm Bondi}$ that is several orders of magnitude higher than other galaxies that have been interpreted to have a radiatively inefficient accretion flow, though it is somewhat smaller than the canonical efficiency of 1--10\%. Taking this at face value, we favor either a standard, efficient, geometrically thin, optically thick disk for the accretion system in Cen~A, or a hybrid model in which a radiatively inefficient optically thin inner accretion flow is surrounded by a standard thin disk (e.g., \citealt{nar96}). We finally note that $\eta_{\rm Bondi}$ for Cen~A and the other sources in Table~\ref{bondi} is an upper limit, as we have ignored any role that cold gas might have in accreting onto the black hole. We also cannot be certain of the true value of $M_{\rm BH}$.

\section{CONCLUSIONS}
We have presented results from one \Ch HETGS and two \XMM EPIC observations of the nucleus of the nearby radio galaxy Centaurus A.  We find that:

\begin{enumerate}

\item Resolved Fe K$\alpha$ line emission at 6.4 keV is detected with \emph{Chandra}, with an rms line width of $\sim$20 $\pm$ 10 eV (90\% confidence). The measured line energy of the Fe K$\alpha$ line, together with its narrowness and lack of response to continuum variations suggests an origin in cool, neutral or near-neutral material far from the central black hole. It is possible that this material takes the form of either an absorber that almost entirely surrounds the central engine or a torus that lies predominantly out of the line of sight. However, we cannot rule out a contribution to the line emission from the outer parts of an accretion disk. Fluorescent K$\alpha$ emission from neutral Si at 1.74 keV and neutral S at 2.30 keV are also detected.

\item The hard (3--10 keV) X-ray spectrum is well-fitted by a heavily
absorbed power-law model with a column density of $\sim$10\(^{23}\)
atoms cm$^{-2}$ and a photon index of $\sim$1.7, consistent with
previous observations of Cen~A. This component is variable by 57\% on
timescales of months but not on shorter timescales of kiloseconds. A
single, heavily absorbed power-law model does not adequately fit the spectrum of Cen~A, as at low energies ($\sim$2 keV) an excess is found. We have discussed two more complex models to explain the observed X-ray emission. In the first model, $\sim$90\% of the nucleus is covered by an absorbing column of  $\sim$10\(^{23}\) atoms cm$^{-2}$, and $\sim$10\% remainder by an absorbing column of $\sim$4$\times$10\(^{22}\) atoms cm$^{-2}$. In the second model, we observe unresolved X-ray emission from two distinct power-law components, each with some intrinsic absorption.

\item The model with a second, less absorbed power-law component is consistent with X-ray emission produced in the subparsec-scale radio jet in Cen~A. We show that the measured X-ray luminosity of the second power law component in Cen~A lies on or near the trendline established for other low-power radio galaxy cores.

\item Using Bondi accretion theory, we estimate the accretion rate of gravitationally captured hot X-ray--emitting gas onto the central black hole to be $\sim$6$\times$10\(^{-4}\) M$_\odot$ yr$^{-1}$. We calculate an efficiency, $\eta_{\rm Bondi}$, by comparing the measured X-ray luminosity with the calculated Bondi luminosity. We find the upper limit of $\eta_{\rm Bondi}$ to be $\sim$0.2\%, which is several orders of magnitude higher than other galaxies thought to have radiatively inefficient accretion flows, though smaller than the canonical efficiency for a standard disk. We favor an accretion model in the form of a standard, geometrically thin, optically thick disk, although note that there are uncertainties associated with the role that cold gas might have in the accretion process, together with the assumed value of $M_{\rm BH}$.

\end{enumerate}

\acknowledgements

We are grateful for support for this work from PPARC (a Studentship
for D.A.E. and research grant for D.M.W.), the Royal Society (Research
Fellowship for M.J.H.), and NASA (contracts NAS8-38248 and NAS8-39073
with the Smithsonian Astrophysical Observatory). D.A.E. thanks the
Harvard-Smithsonian Center for Astrophysics for its hospitality. We
are grateful to the Almudena Prieto and the anonymous referee for useful comments.

\clearpage
\begin{figure}
\plotone{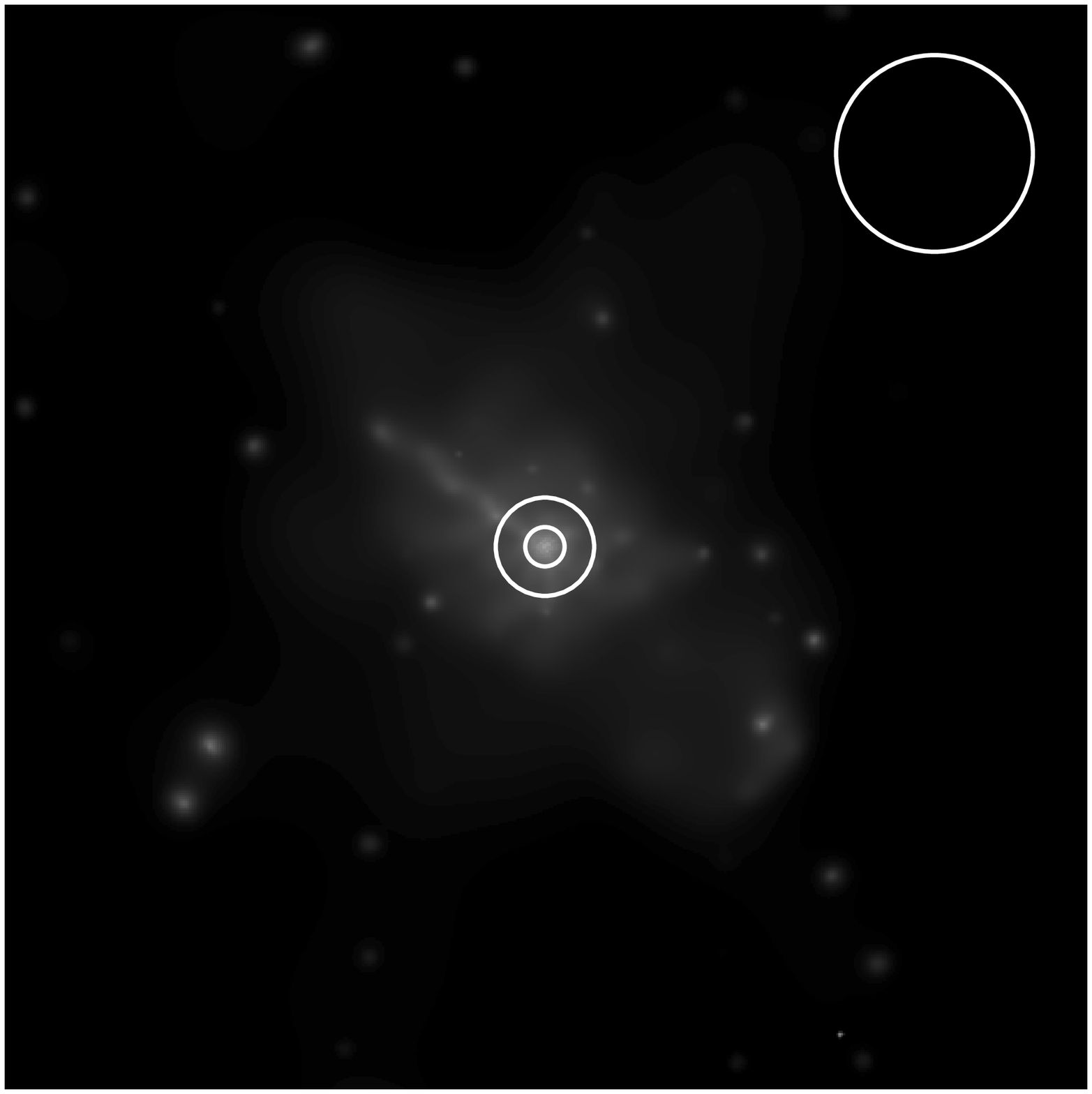}
\caption{Co-added, adaptively smoothed, MOS1, MOS2, and pn image of the second \XMM observation of Cen~A. The energy band is 0.5--4 keV so that both diffuse and pointlike emission may be seen. Pattern selection is 0--4 for the MOS cameras and 0--12 for the pn. No exposure or vignetting correction has been made. A source-centered annulus of inner radius 20$''$ and outer radius 50$''$ is shown, together with a large off-source circle used for local background.}\label{xmmcombined}
\end{figure}

\clearpage
\begin{figure}
\plotone{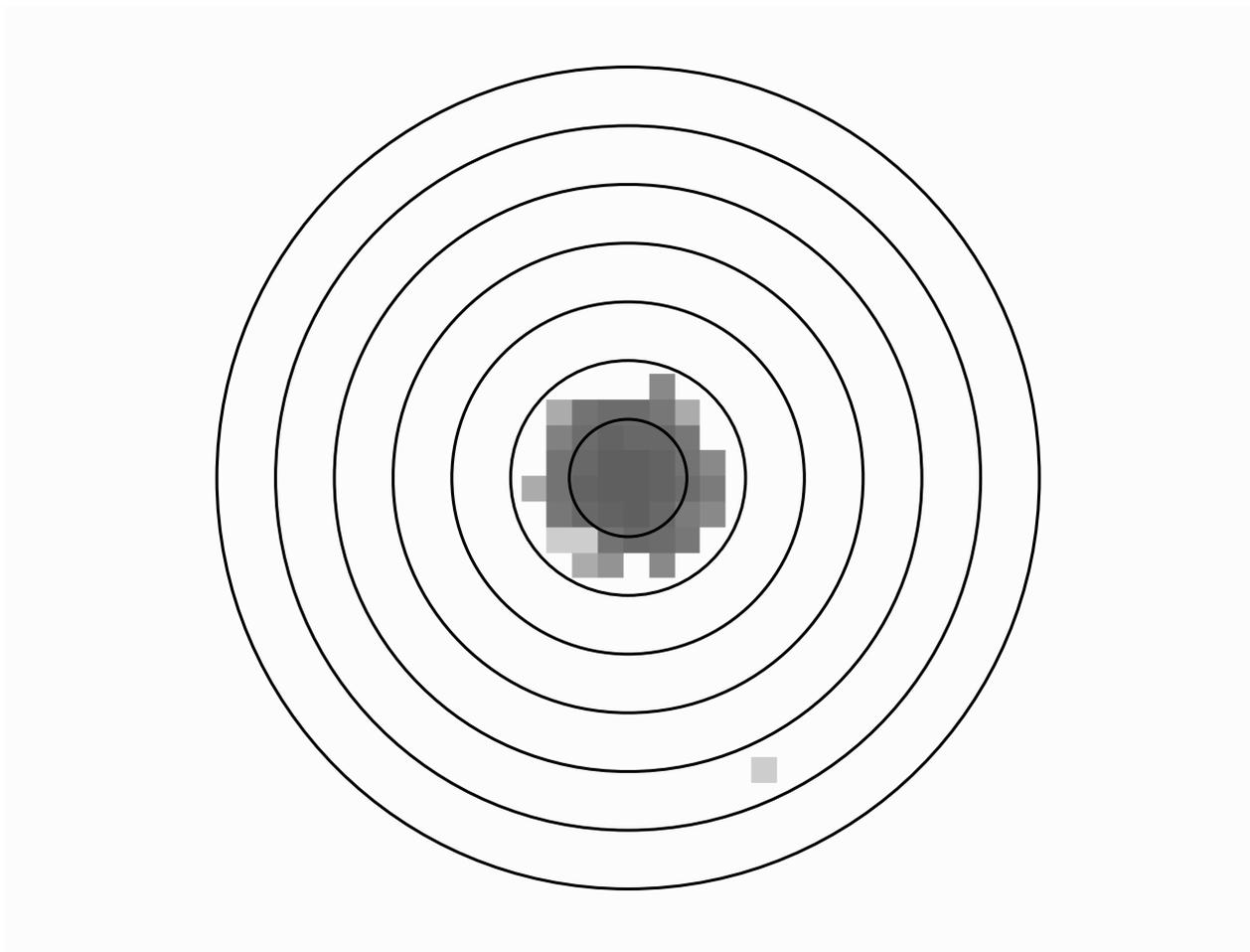}
\caption{\XMM pn image comprised of events that cannot be produced by
a single photon, i.e. those predominantly produced by pileup. This is
produced by selecting events with the \XMM FLAG XMMEA\_0. Concentric
source-centered circles of radii 10$''$--70$''$ are shown. This shows
that pileup is insignificant at distances greater than 20$''$ from the center, and so motivates our choice of extraction region.}\label{pn_xmmea0}
\end{figure}

\clearpage
\begin{figure}
\epsscale{0.7}
\plotone{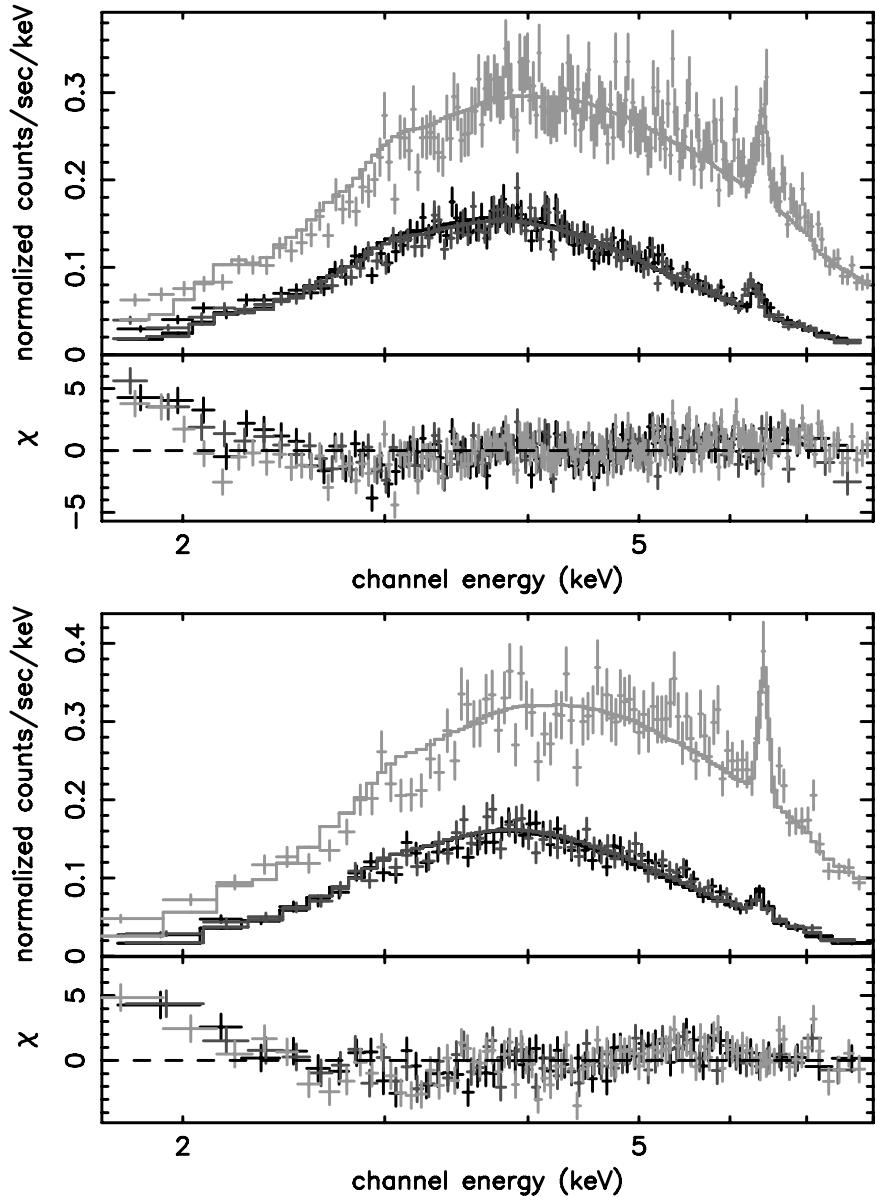}
\caption{Spectral fit to data from the \XMM observations with a single absorbed power-law model, with contributions to $\chi$\(^{2}\) shown. The top panel shows data from the first \XMM observation, and the bottom panel data from the second \XMM observation. The data from the MOS1, MOS2, and pn cameras are denoted, respectively, in black, dark gray, and light gray. This shows the poor fit at low energies and the need for a more complex spectral model.}\label{xmm_1pl}
\end{figure}

\clearpage
\begin{figure}
\epsscale{1.0}
\plotone{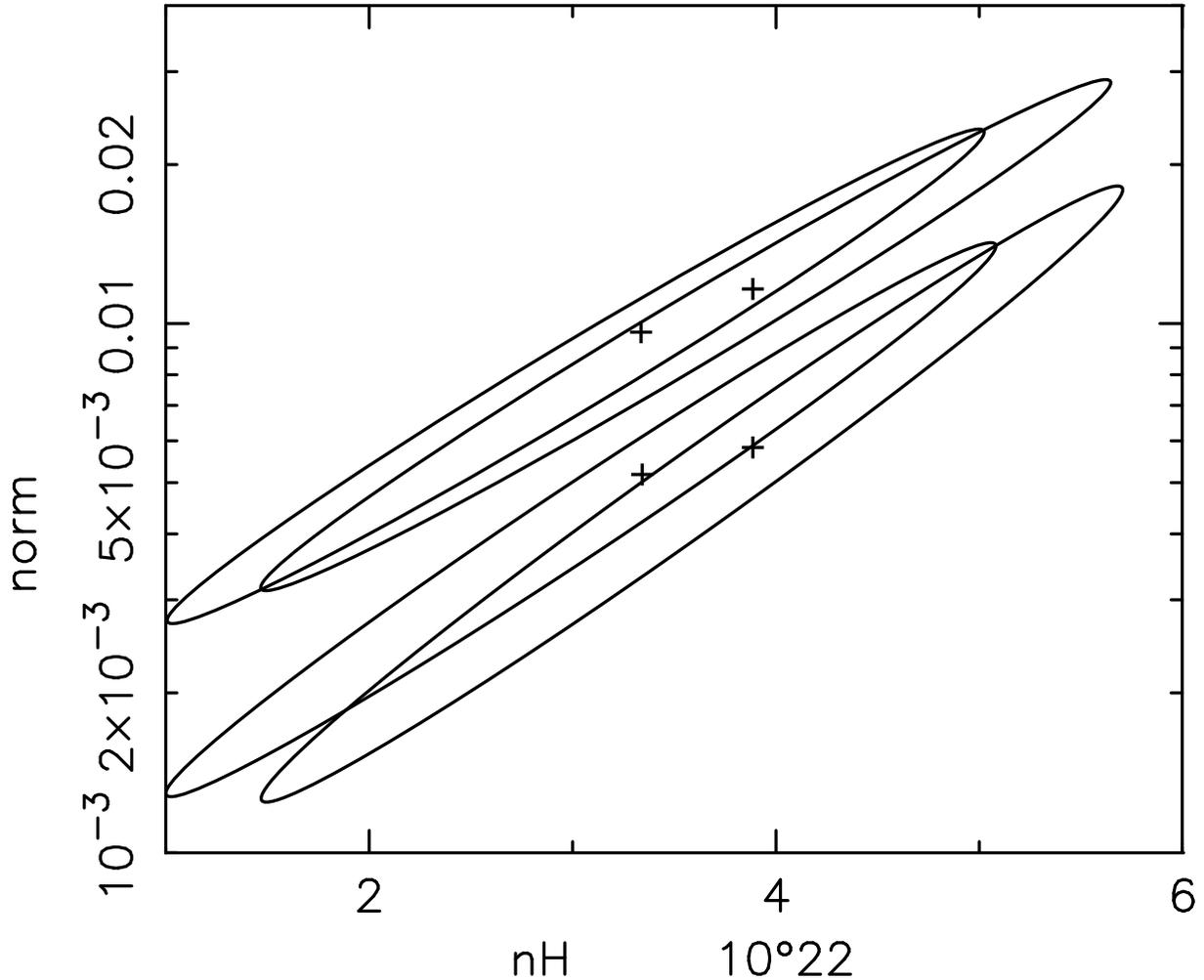}
\caption{\XMM first observation, second power-law 90\% confidence contours ($\chi$\(^{2}\)$_{\rm min}$ + 4.61) for MOS1 ({\it{top two}}) and pn ({\it{bottom two}}). The confidence contours shown for each camera correspond to 90\% confidence upper and lower flux limits for the contribution of jet and gas emission in the annulus, as modeled using the \Ch ACIS CCD's. The disagreement in normalization between the MOS and pn cameras is discussed in the text.}\label{eee_obs1_90}
\end{figure}

\clearpage
\begin{figure}
\epsscale{0.7}
\plotone{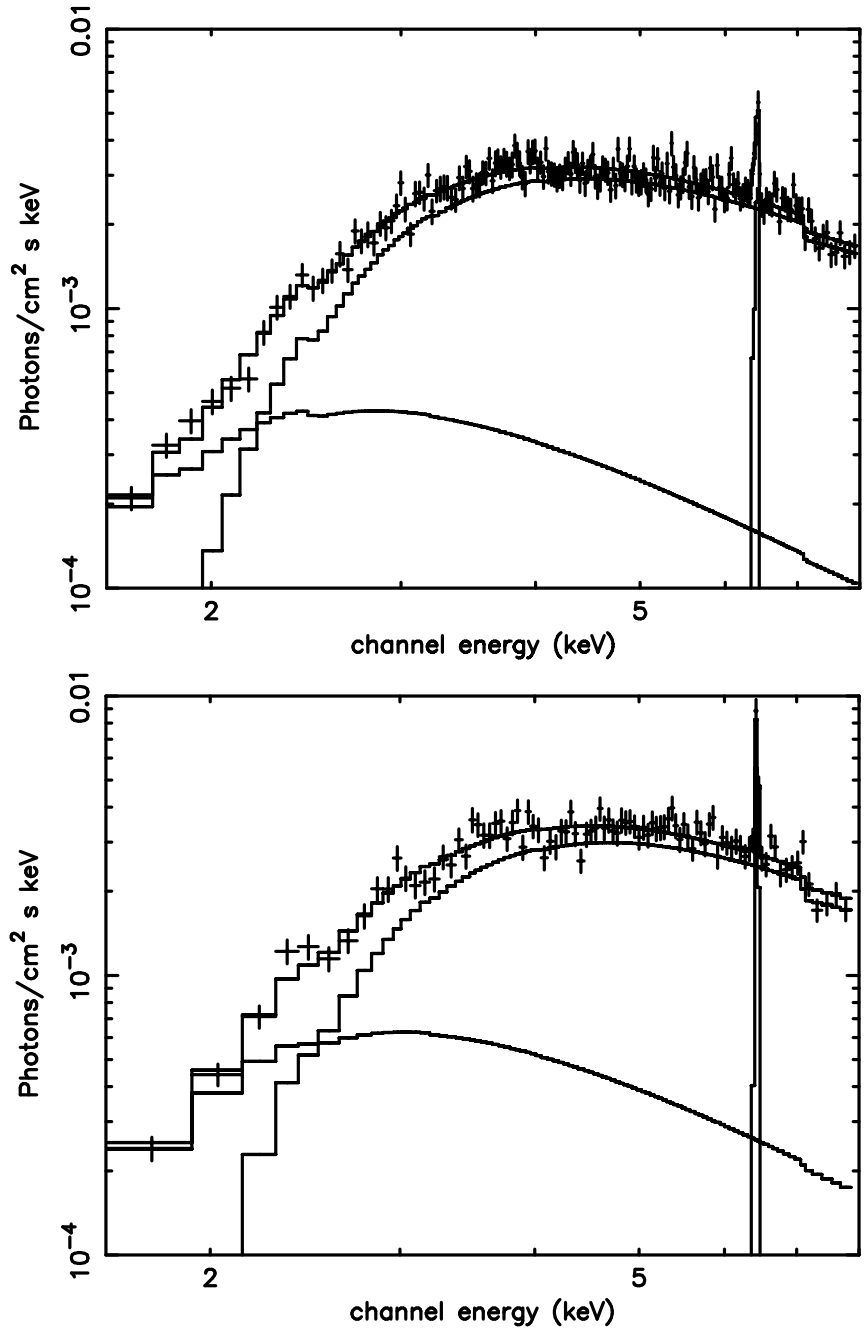}
\caption{Unfolded spectrum for the two \XMM observations showing the
model fit to two absorbed power laws and an Fe K$\alpha$ line. The upper panel shows data from the first \XMM observation, and the lower panel data from the second \XMM observation. For clarity, only the pn data are shown.}\label{xmm_2pl_ufs}
\end{figure}

\clearpage
\begin{figure}
\epsscale{1.0}
\plotone{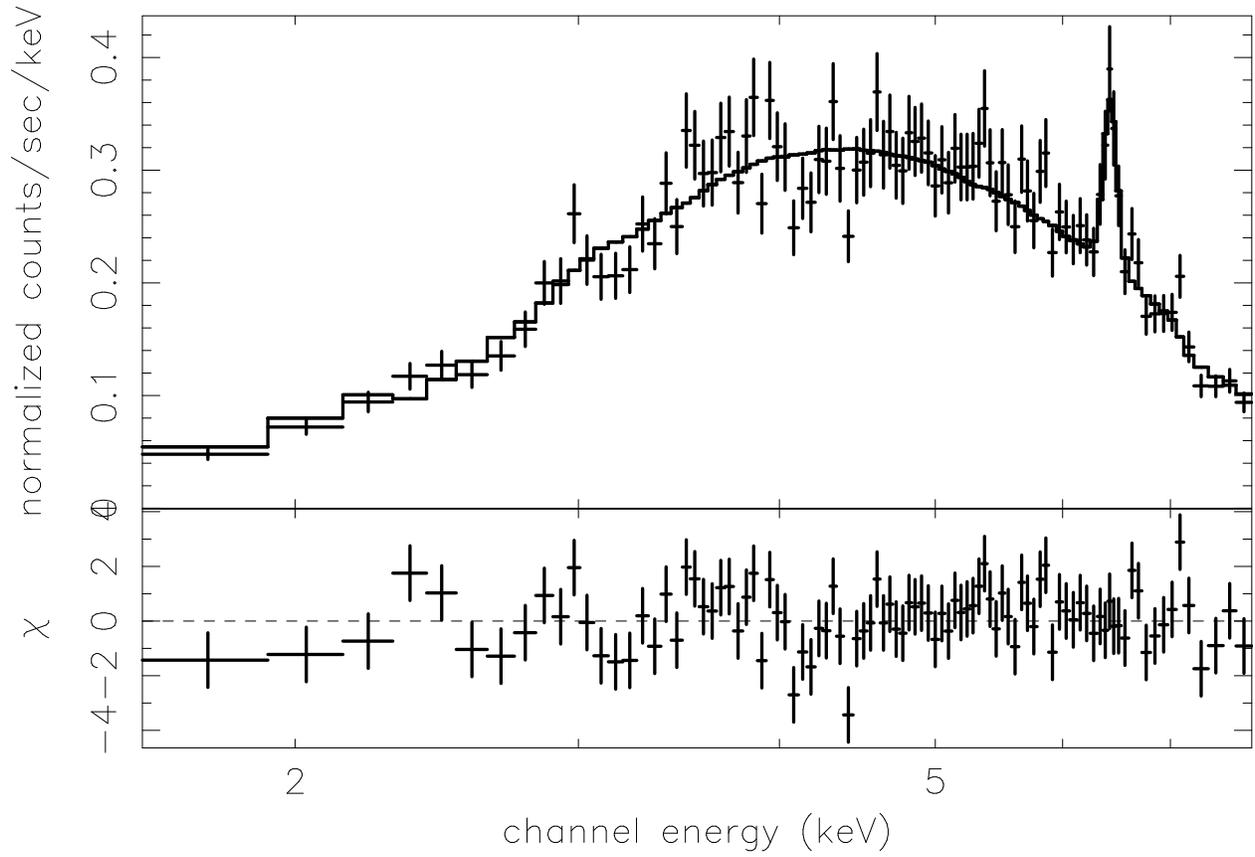}
\caption{The pn counts spectrum for the second \XMM observation showing the model fit to a partially-covered absorber with one power-law and an Fe K$\alpha$ line. This model provides an adequate description of the observed low energy emission.}\label{xmm_zpc_obs2}
\end{figure}

\clearpage
\begin{figure}
\plotone{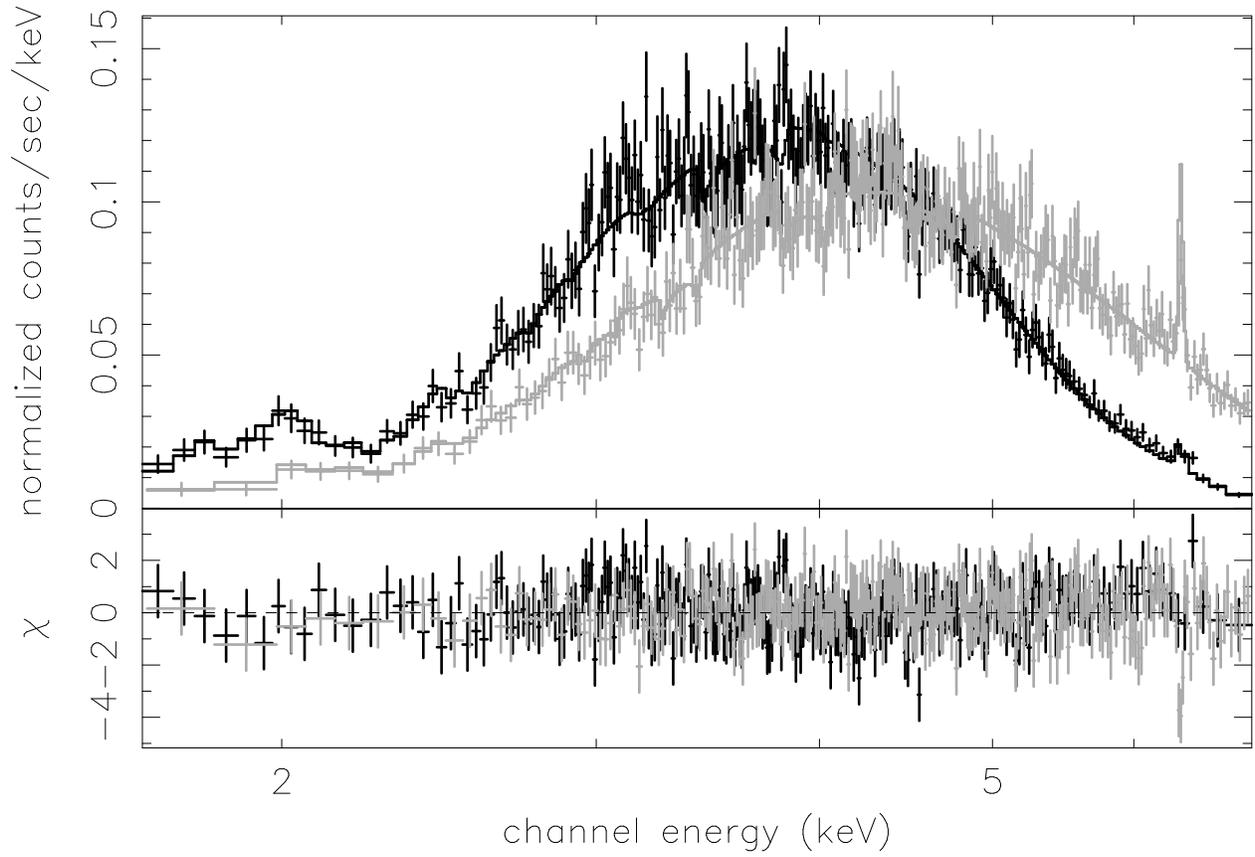}
\caption{\Ch counts spectrum showing the model fit to two absorbed power-laws and an Fe K$\alpha$ line. The MEG-1 data and model are denoted by in black and the HEG-1 data and model in light gray.}\label{chandra_2pl_counts}
\end{figure}

\clearpage
\begin{figure}
\plotone{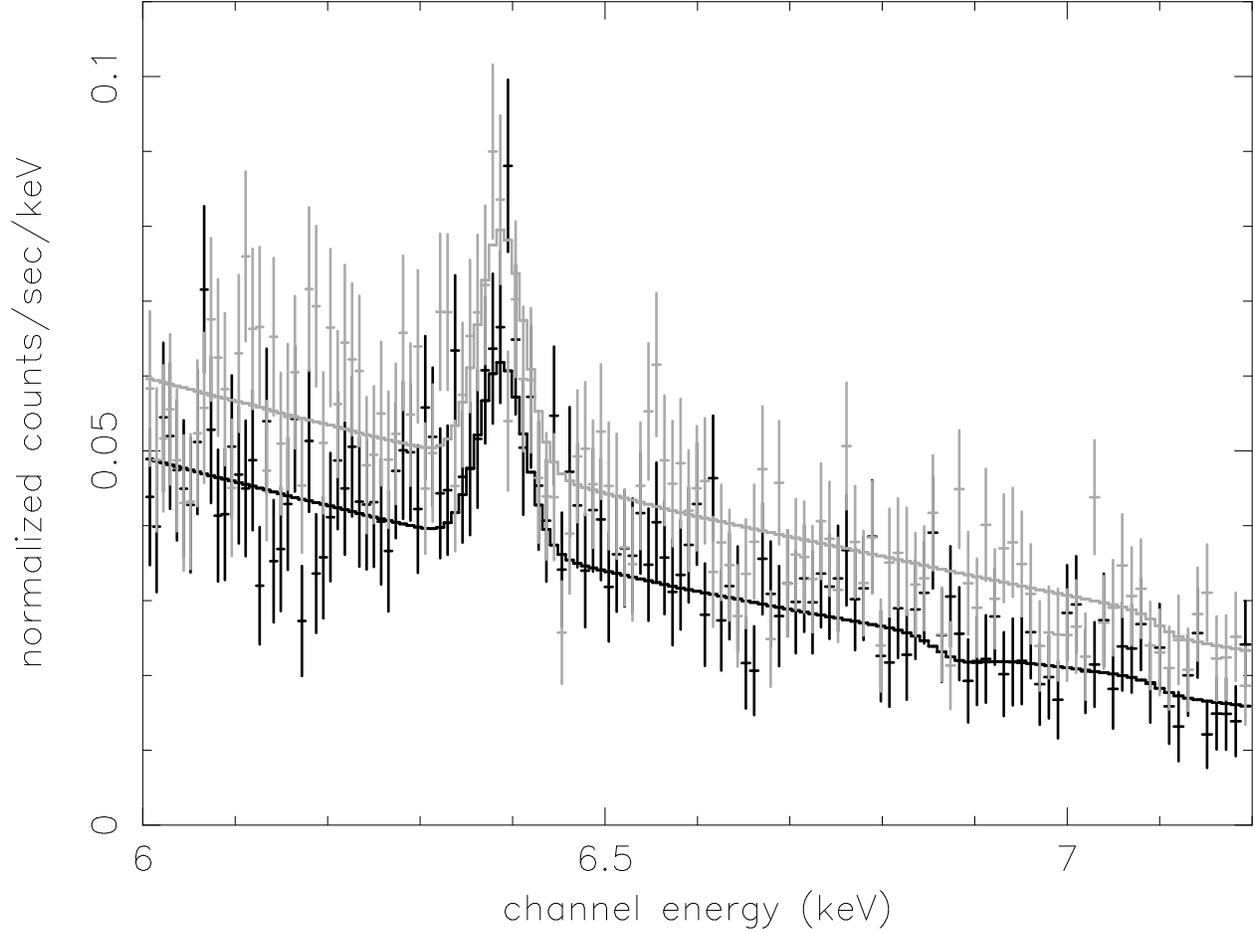}
\caption{The 6.0--7.2 keV counts spectrum for the \Ch HEG+1 ({\it{black}}) and HEG-1 ({\it{light gray}}) observations, showing the model fit to Fe K$\alpha$ line emission, modelled as a doublet, and the nuclear continuum. The Fe K$\alpha$ lines were found to be broadened, with a line width $\sim$20 eV.}
\label{chandra_iron_counts}
\end{figure}

\clearpage
\begin{figure}
\plotone{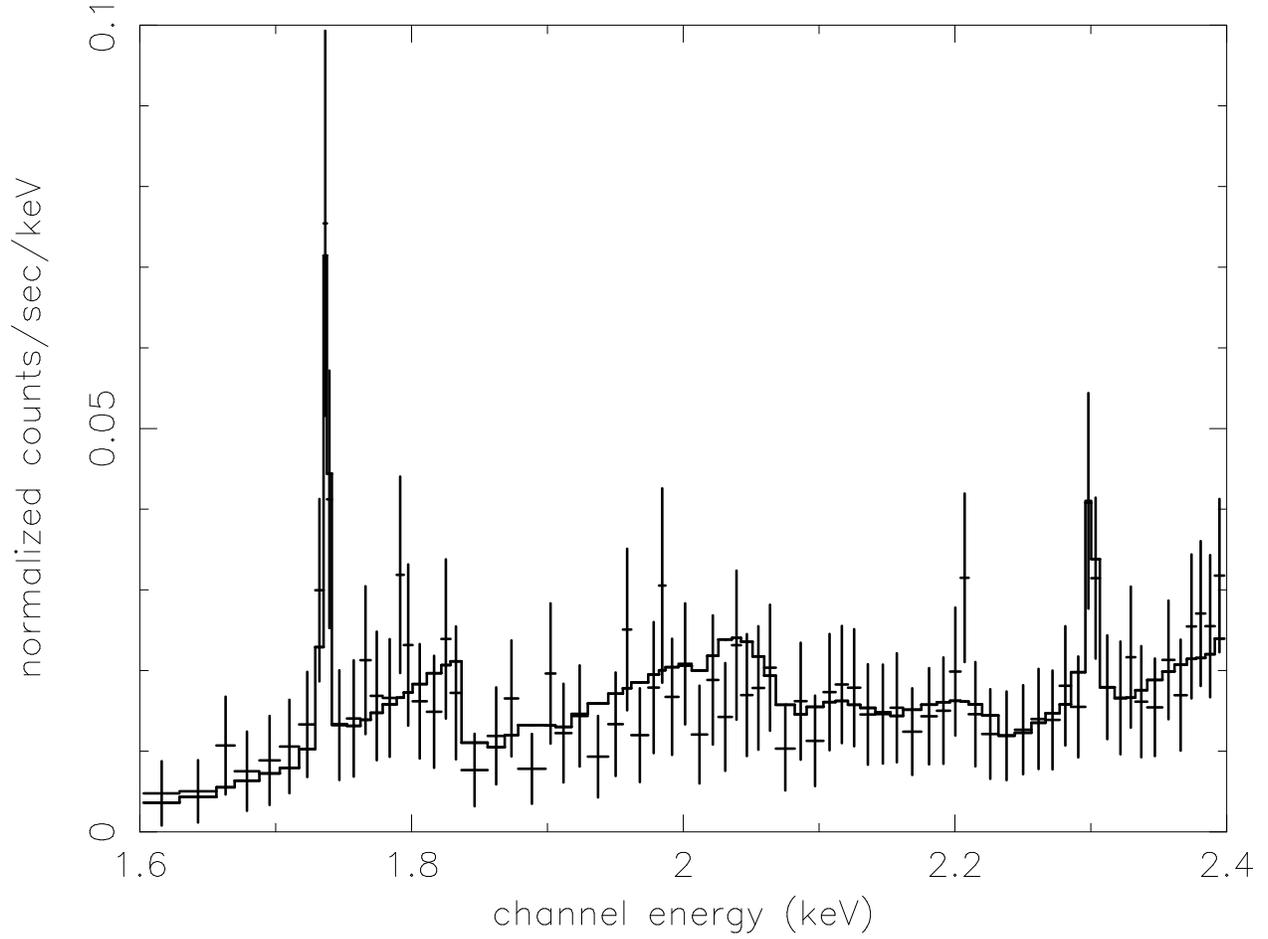}
\caption{The 1.6--2.4 keV counts spectrum for the \Ch MEG-1 observation, showing unresolved K$\alpha$ emission lines from neutral Si at 1.74 keV, neutral S at 2.30 keV, and the nuclear continuum.}
\label{chandra_Si_S}
\end{figure}

\clearpage
\begin{figure}
\plotone{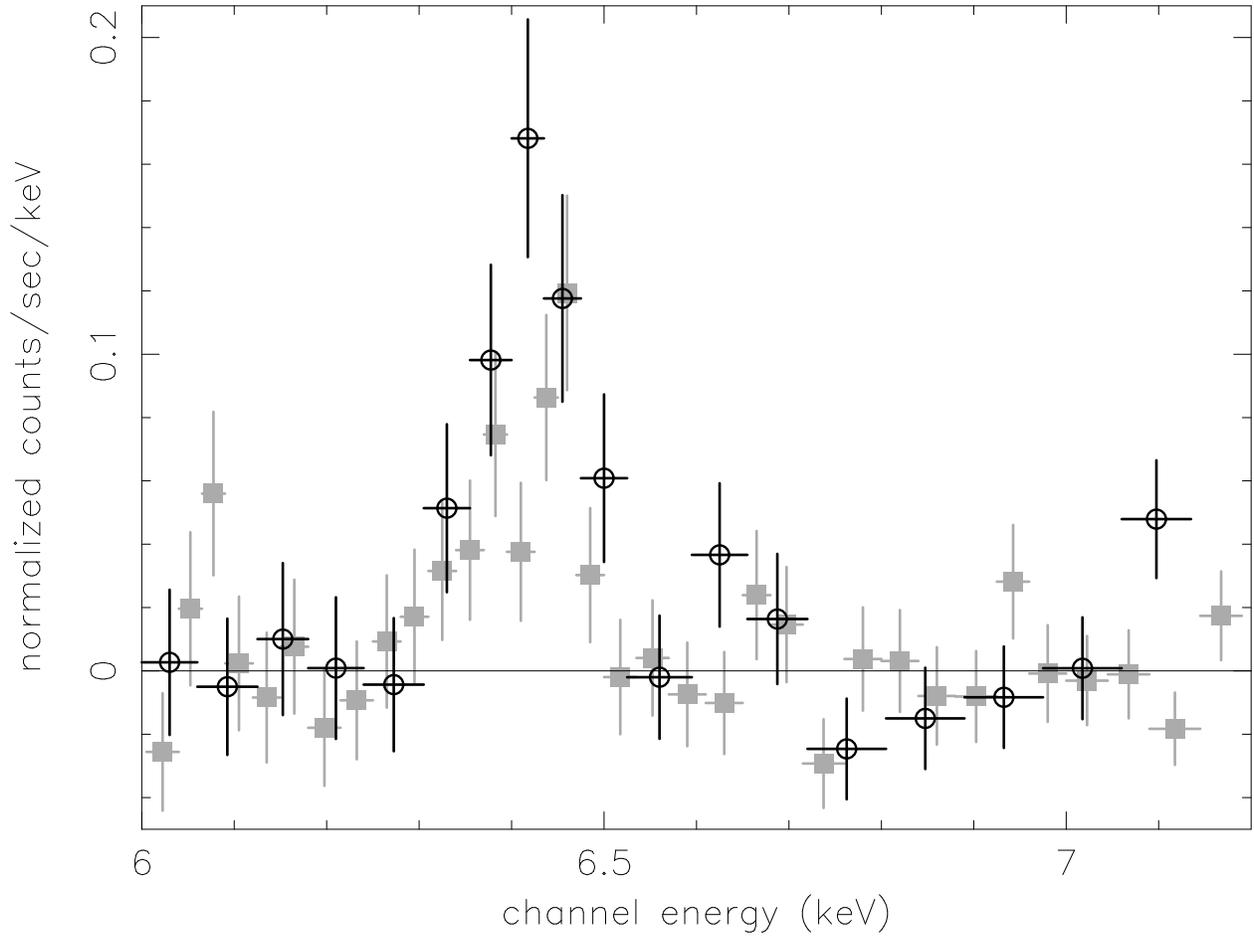}
\caption{6.0--7.2 keV residuals for the \XMM pn observations, after
fitting the nuclear spectrum to a hard, heavily-absorbed
power-law. The first observation is represented by gray squares and
the second by black circles. Emission from Fe K$\alpha$ at 6.4 keV is clearly apparent.}
\label{xmm_iron_residuals}
\end{figure}

\clearpage
\begin{figure}
\plotone{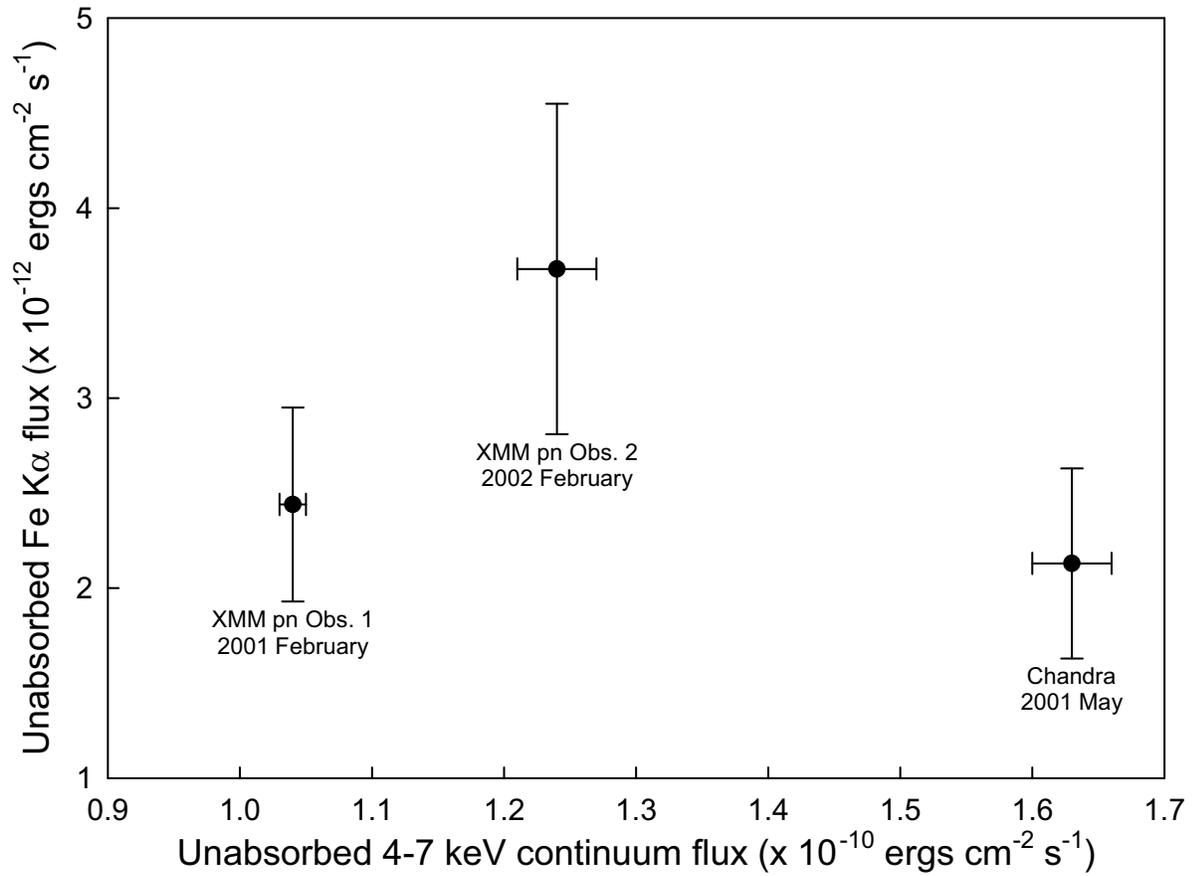}
\caption{Unabsorbed Fe K$\alpha$ line flux vs. unabsorbed 4--7 keV continuum flux for the two \XMM observations (pn camera only), and the \Ch HETGS observation, with 90\% confidence errors shown. This shows the lack of response of the Fe K$\alpha$ line to changes in the continuum flux.}
\label{iron_cont_corr}
\end{figure}

\clearpage
\begin{figure}
\epsscale{0.7}
\plotone{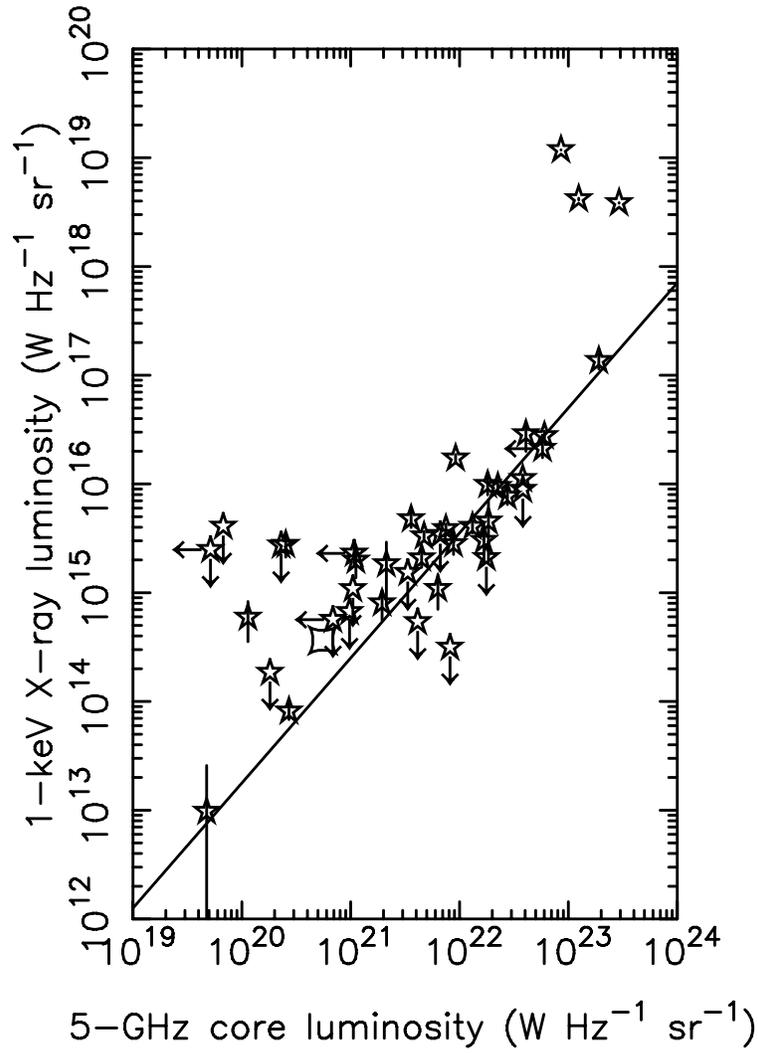}
\caption{Soft X-ray luminosity (as measured with {\it{ROSAT}}) plotted
against 5 GHz radio core luminosity for the B2 sample of radio
galaxies ({\it{stars}}; see \citealt{can99} for a full explanation),
and Cen~A ({\it{squashed box}}). This shows that the Cen~A lies near
the previously established trendline (solid line) for the B2 galaxies.}\label{b2_cena_corr_lumin}
\end{figure}

\clearpage
\begin{deluxetable}{llcc}
\tablecaption{\XMM and \Ch best fit continuum spectral parameters with a two power-law model}
\tabletypesize{\scriptsize}
\tablehead{&&&1 keV normalization\\Observation & Component & Parameters & (photons cm$^{-2}$s$^{-1}$keV$^{-1}$)}
\startdata
      \XMM 1        & 1st Power Law & \nh = (1.19 $\pm$ 0.13)$\times$10\(^{23}\) atoms cm$^{-2}$           & MOS1 = (9.36 $\pm$ 1.51)$\times$10\(^{-2}\) \\ 
                    &               & $\Gamma$ = 1.74$^{+0.11}_{-0.09}$                                    & MOS2 = (8.39 $\pm$ 1.34)$\times$10\(^{-2}\) \\
                    &               &                      \nodata                                         & pn = (7.32 $\pm$ 1.25)$\times$10\(^{-2}\) \\ 
                    & 2nd Power Law & \nh $\simeq$ (3.6$^{+2.2}_{-2.3})$$\times$10\(^{22}\) atoms cm$^{-2}$  & MOS1 = (9.52$^{+13.40}_{-6.63}$)$\times$10\(^{-3}\) \\
                    &               & $\Gamma$ = 2 (fixed)                                                 & MOS2 = (9.98$^{+13.74}_{-6.96}$)$\times$10\(^{-3}\) \\
                    &               &                      \nodata                                         & pn = (5.40$^{+7.66}_{-4.17}$)$\times$10\(^{-3}\) \\
     \Ch HETGS      & 1st Power Law & \nh = (1.00$\pm$ 0.06)$\times$10\(^{23}\) atoms cm$^{-2}$            & (9.65 $\pm$ 1.41)$\times$10\(^{-2}\)\\         
                    &               & $\Gamma$ = 1.64$\pm$0.07                                             & \nodata \\
                    & 2nd Power Law & \nh = 3.8 $\times$10\(^{22}\) atoms cm$^{-2}$ (fixed)                & (1.74$^{+1.46}_{-1.45}$)$\times$10\(^{-3}\)       \\
                    &               & $\Gamma$ = 2 (fixed)                                                 & \nodata \\
      \XMM 2        & 1st Power Law & \nh = (1.40 $\pm$ 0.20)$\times$10\(^{23}\) atoms cm$^{-2}$           & MOS1 = (10.39 $\pm$ 3.13)$\times$10\(^{-2}\) \\ 
                    &               & $\Gamma$ = 1.77$^{+0.19}_{-0.15}$                                    & MOS2 = (9.79 $\pm$ 2.90)$\times$10\(^{-2}\) \\
                    &               &                      \nodata                                         & pn = (8.35 $\pm$ 2.60)$\times$10\(^{-2}\) \\ 
                    & 2nd Power Law & \nh $\simeq$ (4.0$^{+1.8}_{-1.6})$$\times$10\(^{22}\) atoms cm$^{-2}$  & MOS1 = (11.10$^{+8.51}_{-5.12}$)$\times$10\(^{-3}\) \\
                    &               & $\Gamma$ = 2 (fixed)                                                 & MOS2 = (11.29$^{+8.81}_{-5.33}$)$\times$10\(^{-3}\) \\
                    &               &                      \nodata                                         & pn = (6.04$^{+6.52}_{-3.18}$)$\times$10\(^{-3}\) \\

\enddata
\label{2pow_fitted}
\tablecomments{Uncertainties are 90\% for one interesting parameter (i.e., $\chi$\(^{2}\)$_{\rm min}$ + 2.7).}
\end{deluxetable}

\clearpage
\begin{deluxetable}{lcc}
\tablecaption{\XMM best fit continuum spectral parameters for a partial covering model}
\tabletypesize{\scriptsize}
\tablehead{&&1 keV normalization\\Observation & Joint parameters &(photons cm$^{-2}$s$^{-1}$keV$^{-1}$)}
\startdata
      \XMM 1         & $N_{\rm H, 1}$ = (4.28$^{+0.93}_{-1.18}$)$\times$10\(^{22}\) atoms cm$^{-2}$  & MOS1 = (9.67$^{+2.10}_{-1.75}$)$\times$10\(^{-2}\) \\ 
                    & $N_{\rm H, 2}$ = (7.53$^{+0.65}_{-0.57}$)$\times$10\(^{22}\) atoms cm$^{-2}$  & MOS2 = (8.92$^{+1.93}_{-1.61}$)$\times$10\(^{-2}\) \\
                    & $\Gamma$ = 1.72 $\pm$ 0.10                                                    & pn =   (7.21$^{+1.56}_{-1.32}$)$\times$10\(^{-2}\) \\ 
                    & Covering fraction = 0.89$^{+0.05}_{-0.07}$                                    & \nodata                                          \\
      \XMM 2         & $N_{\rm H, 1}$ = (4.54$^{+0.65}_{-1.75}$)$\times$10\(^{22}\) atoms cm$^{-2}$  & MOS1 = (10.73$^{+4.40}_{-3.04}$)$\times$10\(^{-2}\) \\ 
                    & $N_{\rm H, 2}$ = (8.81$^{+1.48}_{-0.90}$)$\times$10\(^{22}\) atoms cm$^{-2}$  & MOS2 = (10.30$^{+4.21}_{-2.91}$)$\times$10\(^{-2}\) \\
                    & $\Gamma$ = 1.73 $\pm$ 0.17                                                    & pn =   (8.34$^{+3.42}_{-2.37}$)$\times$10\(^{-2}\) \\ 
                    & Covering fraction = 0.89$^{+0.05}_{-0.07}$                                    & \nodata                                          \\
\enddata
\label{xmm_zpc_obs2_table}
\tablecomments{Uncertainties are 90\% for one interesting parameter (i.e., $\chi$\(^{2}\)$_{\rm min}$ + 2.7).}
\end{deluxetable}
\clearpage

\clearpage
\begin{deluxetable}{llcccc}
\tablecaption{Emission line parameters}
\tabletypesize{\scriptsize}
\tablehead{&&&&&Unabsorbed flux\\Line & Observation & Centroid (keV) & Line width (eV) & Equivalent width (eV) & ($\times$10\(^{-12}\) ergs cm$^{-2}$ s$^{-1}$)}
\startdata
Fe K$\alpha$ & \XMM 1       & 6.417$^{+0.034}_{-0.009}$         & 10 (fixed)             & 81.6$^{+33.4}_{-33}$    & 2.44$^{+0.52}_{-0.49}$   \\
             & \Ch HETGS   & K$\alpha$$_1$ = 6.404 $\pm$ 0.006 & 20$^{+8.99}_{-8.11}$   & 43.6$^{+18.0}_{-18.7}$  & 2.13 $\pm$ 0.50          \\
             &             & K$\alpha$$_2$ = 6.398             & \nodata                & \nodata                 & \nodata                       \\
             & \XMM 2       & 6.428$^{+0.027}_{-0.015}$         & 10 (fixed)             & 99.9$^{+35.1}_{-52.9}$  & 3.68$^{+0.87}_{-0.86}$   \\
Si K$\alpha$ & \Ch HETGS   & 1.740 $\pm$ 0.014                 & 10 (fixed)             & 32.7$^{+6.7}_{-4.1}$    & 2.42 $\pm$ 0.97          \\
S  K$\alpha$ & \Ch HETGS   & 2.304 $\pm$ 0.024                 & 10 (fixed)             & $<$ 10                  & $<$ 0.77                 \\
\enddata
\label{line_parameters}
\tablecomments{The \XMM line widths for the Fe K$\alpha$ line, and the \Ch line widths for the Si K$\alpha$ line and S K$\alpha$ lines were fixed at 10 eV (i.e., unresolved). The equivalent widths were measured relative to the primary continuum. Uncertainties are 90\% for one interesting parameter (i.e., $\chi$\(^{2}\)$_{\rm min}$ + 2.7).}
\end{deluxetable}

\clearpage
\begin{deluxetable}{lccc}
\tablecaption{\XMM and \Ch absorbed and unabsorbed 4--7 keV fluxes, and 2--10 keV unabsorbed luminosities}
\tabletypesize{\scriptsize}
\tablehead{& 4--7 keV absorbed flux & 4--7 keV unabsorbed flux & 2--10 keV unabsorbed luminosity \\Observation & ($\times$10\(^{-10}\) ergs cm$^{-2}$ s$^{-1}$) & ($\times$10\(^{-10}\) ergs cm$^{-2}$ s$^{-1}$) & ($\times$10\(^{41}\) ergs s$^{-1}$)}
\startdata
\XMM pn 1 (2001 February)  & 0.75 $\pm$ 0.01 & 1.04 $\pm$ 0.01 & 3.88 $\pm$ 0.07 \\ 
\Ch HEG+1 (2001 May)       & 1.19 $\pm$ 0.03 & 1.63 $\pm$ 0.03 & 6.25 $\pm$ 0.19 \\ 
\XMM pn 2 (2002 February)  & 0.84 $\pm$ 0.02 & 1.24 $\pm$ 0.03 & 4.28 $\pm$ 0.12 \\
\enddata
\label{4to7var}
\tablecomments{Uncertainties are 90\% for one interesting parameter (i.e., $\chi$\(^{2}\)$_{\rm min}$ + 2.7).}
\end{deluxetable}

\clearpage
\begin{deluxetable}{lccc}
\tablecaption{X-ray and radio flux densities and ratios for the kiloparsec-scale and parsec-scale jets}
\tabletypesize{\scriptsize}
\tablehead{Component & 1 keV X-ray flux density ($\mu$Jy) & Radio flux density (Jy) (frequency) & Ratio ($\times$10\(^{-7}\))}
\startdata
kiloparsec-scale jet                 & 0.22 $\pm$ 0.01                    & 0.74 $\pm$ 0.12 (8.4 GHz)           & 3.0 $\pm$ 0.6                \\
2nd power law / parsec-scale jet  & 3.31                               & 5 (4.8 GHz)                         & 6.6                          \\
\enddata
\label{fluxratios}
\tablecomments{Uncertainties are 90\% for one interesting parameter (i.e., $\chi$\(^{2}\)$_{\rm min}$ + 2.7).}
\end{deluxetable}

\clearpage
\begin{deluxetable}{lcccccccccl}
\tablecaption{Calculation of efficiencies, $\eta$$_{Bondi}$, by comparing the measured X-ray luminosity with the expected Bondi luminosity}
\tabletypesize{\tiny}
\tablehead{& Distance & $M_{\rm SMBH}$ & $\dot{M}_{\rm Bondi}$ & $L_{\rm Edd}$ & $\dot{M}_{\rm Bondi} c$\(^{2}\) & $L_{\rm X}$ & Energy range & &&\\ Galaxy & (Mpc) & (M$_\odot$) & (M$_\odot$ yr\(^{-1}\)) & (ergs s\(^{-1}\)) & (ergs s\(^{-1}\)) & (ergs s\(^{-1}\)) & (keV) & $\eta_{\rm Bondi}$ & Ref & Interpretation}
\startdata
Milky Way  & 8.0$\times$10\(^{-3}\) & 2.6$\times$10\(^{6}\) & 1.0$\times$10\(^{-5}\) & 3.3$\times$10\(^{44}\) & 1.0$\times$10\(^{42}\) & 2.4$\times$10\(^{33}\)       & 2--10   & 2.4$\times$10\(^{-9}\) & 1,2 & Inefficient \\
Cen A      & 3.4                    & 2.0$\times$10\(^{8}\) & 6.4$\times$10\(^{-4}\) & 2.6$\times$10\(^{46}\) & 2.1$\times$10\(^{44}\) & 4.8$\times$10\(^{41}\)       & 2--10   & 2.3$\times$10\(^{-3}\) & 3   & Standard    \\
NGC 4636   & 15.0                   & 7.9$\times$10\(^{7}\) & 8.0$\times$10\(^{-5}\) & 1.0$\times$10\(^{46}\) & 4.5$\times$10\(^{42}\) & 2.7$\times$10\(^{38}\)       & 2--10   & 6.0$\times$10\(^{-5}\) & 4   & Inefficient \\
NGC 4472   & 16.7                   & 5.7$\times$10\(^{8}\) & 7.9$\times$10\(^{-3}\) & 7.2$\times$10\(^{46}\) & 4.5$\times$10\(^{44}\) & 6.4$\times$10\(^{38}\)       & 2--10   & 1.4$\times$10\(^{-6}\) & 5   & Inefficient \\
M87        & 18.0                   & 3.0$\times$10\(^{9}\) & 1.0$\times$10\(^{-1}\) & 3.8$\times$10\(^{47}\) & 5.0$\times$10\(^{45}\) & 7.0$\times$10\(^{40}\)       & 0.5--7  & 1.4$\times$10\(^{-5}\) & 5   & Inefficient \\
NGC 1399   & 20.5                   & 1.1$\times$10\(^{9}\) & 4.0$\times$10\(^{-2}\) & 1.4$\times$10\(^{47}\) & 2.3$\times$10\(^{45}\) & 9.7$\times$10\(^{38}\)       & 2--10   & 4.2$\times$10\(^{-7}\) & 4   & Inefficient \\
NGC 4261   & 31.6                   & 4.9$\times$10\(^{8}\) & 4.5$\times$10\(^{-2}\) & 6.2$\times$10\(^{46}\) & 2.5$\times$10\(^{45}\) & 1.2$\times$10\(^{41}\)       & 0.3--10 & 4.8$\times$10\(^{-5}\) & 6   & Inefficient \\
NGC 6251   & 94.4                   & 6.0$\times$10\(^{8}\) & 8.0$\times$10\(^{-3}\) & 7.6$\times$10\(^{46}\) & 4.6$\times$10\(^{44}\) & 8.8$\times$10\(^{42}\)       & 0.4--10 & 1.9$\times$10\(^{-2}\) & 7   & Standard    \\
\enddata
\label{bondi}
\tablerefs{(1) \citealt{qua03}; (2) \citealt{bag03}; (3) Present work; (4) \citealt{loe01}; (5) \citealt{dimatt03}; (6) \citealt{gli03}; (7) \citealt{gli04}}
\end{deluxetable}

\end{document}